# Flood Risk Assessment of the National Harbor at Maryland, United States

Neftalem Negussie[1], Addis Yesserie[1], Chinchu Harris[1], Abou Keita[1], Huthaifa I. Ashqar[1,2]

[1] University of Maryland, Baltimore County

[2] Arab American University

**Abstract**

Over the past few decades, floods have become one of the costliest natural hazards and losses have sharply escalated. Floods are an increasing problem in urban areas due to increased residential settlement along the coastline and climate change is a contributing factor to this increased frequency. In order to analyze flood risk, a model is proposed to identify the factors associated with increased flooding at a local scale. The study area includes National Harbor, MD, and the surrounding area of Fort Washington. The objective is to assess flood risk due to an increase in sea level rise for the study area of interest. The study demonstrated that coastal flood risk increased with sea level rise even though the predicted level of impact is fairly insignificant for the study area. The level of impact from increased flooding is highly dependent on the location of the properties and other topographic information.

## Introduction

Coastal floods are among the most damaging natural disasters in the United States (Gall et al., 2011; Smith and Katz, 2013). Flooding is a major hazard to urban infrastructure; all 50 states have experienced floods or flash floods in the past 5 years (FEMA, 2019). Climate change is not the only contributing factor in changing precipitation and flood patterns (Sharma et al., 2018). However, climate change will increase the likelihood of future severe losses in flood prone areas with increased population and economic growth (Hemmati et al., 2020).

Accessibility to ports, recreational facilities, and fertile agricultural lands have made floodplains and coastal areas desirable places to live, consequently, flood-prone areas are experiencing steady population growth and economic development, leading to increased flood-related risks (Hemmati et al., 2020). The encroachment of urban growth in flood-prone areas, driven by socioeconomic development, has received less attention in comparison to climate change as a source of increased flood risk (Hemmati et al., 2020). Flooding due to sea level rise and coastal storms may increase the likelihood of endangerment and/or damage to life and property as a consequence of inundation and shoreline erosion. The combination of average annual direct and indirect flood losses is nearly $ 8 billion, as reported by the National Oceanic and Atmospheric Administration (NOAA) (2018).

Flooding events are expected to increase in frequency and intensity in the coming years due to rising sea levels and more frequent extreme precipitation events (Jonathan et al, 2013). Despite efforts by federal, state, and local governments to manage losses, flood hazards still threaten the lives of millions of people in the United States. Continued coastal development may even put more properties and infrastructures at risk in the coming decades (Kousky and Michel-Kerjan, 2015; Dinan, 2017; Zagorsky, 2017).

According to the Maryland Department of Natural Resources (DNR), the average rate of sea level rise along Maryland's coastline has been 3-4mm/yr or approximately one foot per century. Such rates are nearly twice of the global average 1.8mm/year, a result probably due to substantial land subsidence. Although the extent of future sea-level rise remains uncertain, sea-level rise generally is anticipated to have a range of economic, social, and environmental effects in the U.S. While coastal flooding is likely to increase, some of the economic and social impacts may be avoided through future investments in adaptation measures (Houser et al., 2015). The United States has previously invested billions of dollars to reduce flood risk and will continue to invest further in this initiative. As a result of this investment, many research and government programs are aimed at investigating and presenting improved approaches to save lives and reduce damage and economic losses (Hemmati et al., 2020).

Flood risk is the product of hazard: the chance of flood occurrence; exposure: the subjected population and value of properties to flooding; and vulnerability: the susceptibility of the exposed elements to the hazard (Kron, 2005; Vahid et al., 2020). In this context, flood risk assessments are an essential tool in flood management. It aids urban planners and policymakers in designing effective strategies to reduce the flood risk in vulnerable areas (Vahid et al., 2020). Moreover, the modeling of flood damage is an important component in flood risk assessments which are the basis for risk-oriented flood management, risk mapping, and financial appraisals (Tina et al., 2014).

In past years, many studies have employed various statistical and machine learning methods to improve flood susceptibility mapping (Vahid et al., 2020). However, flood risk modeling and risk assessments in areas that are highly exposed to sea-level rise and flooding has not been extensively studied. Additionally, holistic coastal flood risk assessments and sea level rise at local scales still needs to be further researched. The main objective of this study is to assess flood risk due to sea-level rise and assess the associated increased cost in current property values based on the area of the property as well as its respective lot in the National Harbor area.

As previously stated, this study is distinct from previous studies as it provides policy implications for urban planners and policy makers at local level governments. Moreover, a rapid flood risk assessment approach provides insights to coastal planners and decision makers for improved future planning. Furthermore, such an assessment would help to improve economic estimation for direct flood damage in residential and commercial areas and several agricultural land-use properties.

## Study Area and Data

*Study area:*

Our study area is located in Prince George's County, Maryland. It covers the area identified as National Harbor and Fort Washington (Fig 1). It is located at 38°44′37″N 77°0′37″W (38.743481, -77.010383). It includes land along the Potomac River near the Woodrow Wilson Bridge and just south of Washington, DC. According to the United States Census Bureau, the census-designated place has a total area of 16.57 square miles (42.9km2), of which 13.79 square miles (35.7km2) of land and 2.78 square miles (7.2km2) of water. The Fort Washington community is located west of Maryland Route 210 with an additional area to the east of the highway.

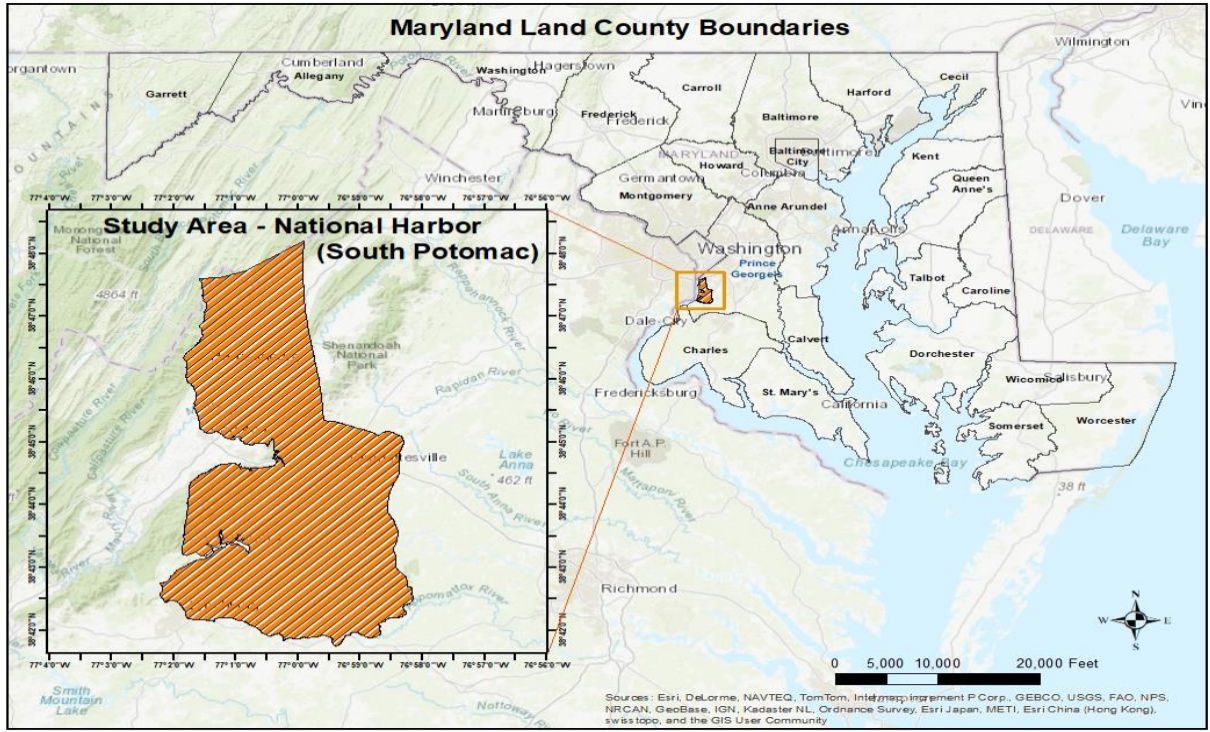

**Figure 1. Location of the Study Area, National Harbor-Fort Washington, Prince George's County, Maryland.**

As of the 2010 census, there were 23,717 individuals, 8,135 households, and 6,319 families residing in the census-designated place (CDP). The population density was 1,719.8 individuals per square mile (664.3/km2). The National Harbor area marks a location of active development within the region. Since this area was rezoned for mixed use in 1994, it has seen an increase in development. It is anticipated that the economic stimulus provided by the new development at the National Harbor will further transform the market in that part of the county, thus incentivizing redevelopments of nearby suburbs such as Oxon Hill.

*Data:*

This study utilized datasets that contain detailed information on property land values and tax assessments, insurance payouts, flood elevation and flood depth for the area of interest. The following datasets were used in this study:

- **Census area boundary:** The South Potomac census block consists of the National Harbor area and was used as the limit of the study area. The boundary was obtained from the Prince George's County Planning GIS repository (https://gisdata.pgplanning.org/metadata/).

- **Study Area Fishnet:** Since the spatial data used in this study was layered, the layers were spliced and then subsequently clustered together. Rectangular grids of 30 m x 30 m (98 ft x 98 ft) were generated encompassing the entire study area.

- **Property Flattened:** This layer is a combination of the property boundary, the Maryland Tax information, and the Land Data file. According to the county's planning department, from which we acquired the dataset, the property line dataset is updated on a daily basis while the other two components (the tax and land data) is updated monthly.

- **Digital Elevation Model (DEM):** This index reflects the topographic change for the area of study. The raster DEM data for the area was obtained from bare earth light detection and ranging (LIDAR) data collected in 2018 for the county and neighboring areas. The DEM has a 2 ft. cell size and was downloaded from the county's GIS repository.

- **Baseflood Elevation (BFE):** This dataset contains the water surface elevation representing the 1% chance of flood discharge to be exceeded in any given year. BFEs are regulatory elevations in the National Flood Insurance Program (NFIP) and are computed for zones with increased flood hazard level which are shown on the Digital Flood Insurance Rate Map (DFIRM). The BFE data was obtained from the DFIRM database that was downloaded from FEMA.

## Methodology

To achieve the goal of developing a flood risk assessment, a Python based approach coupled with linear regression analysis is implemented to process our data. GIS mapping tool has been used to further visualize data (Fig 2).

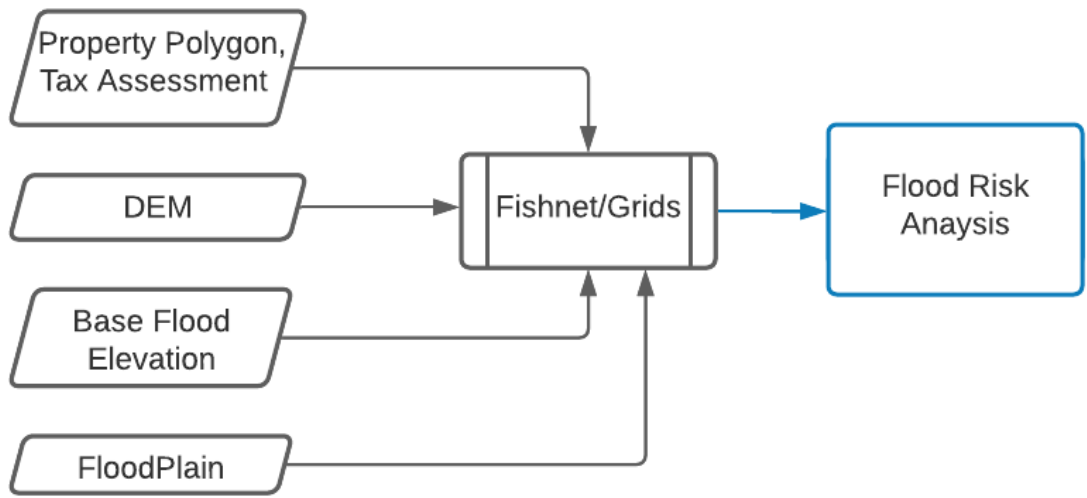

**Figure 2. Flowchart of how the data was compiled in this study using GIS.**

*Spatial analysis:*

- Generate a fishnet grid to cover the study area.
- Intersect the Property Flattened data with the fishnet grid to transfer the area information into individual squares within the fishnet. Since the Property Flattened data contains both the assessed value and the plot size, the cost will have to be weighted using an area weighting method taking the total area of the plot from the dataset and computing how much of the fraction of the area is within the fishnet grid.
- Assign elevation for each area in the fishnet grid from the DEM using the zonal statistics function in the GIS. This will assign average elevation for the individual areas in the grid.

- Assign the base flood information obtained from the BFE data into the grids and compute the flooding depth as the difference between the ground elevation and the base flood elevation. Depth values that are computed negative will mean that the ground elevation is higher than the flood elevation and therefore above flooding (Fig 3).
- For every fishnet area with a positive flooding depth, compute the damage cost using the cost of flooding model developed by FEMA. The maximum the cost will go for each area will be limited to the initial cost generated from the Property Flattened dataset since it should not cost more than what the property is assessed *(https://www.floodsmart.gov/flood-insurance-cost/calculator)*.

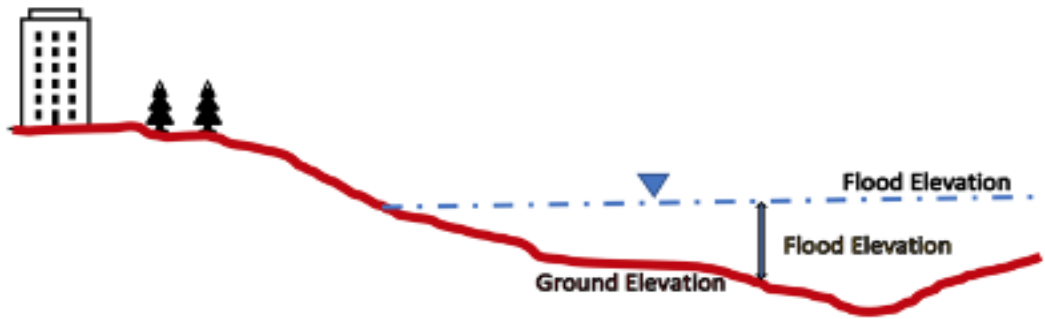

**Figure 3. Description of Flood depth and Flood Elevation.**

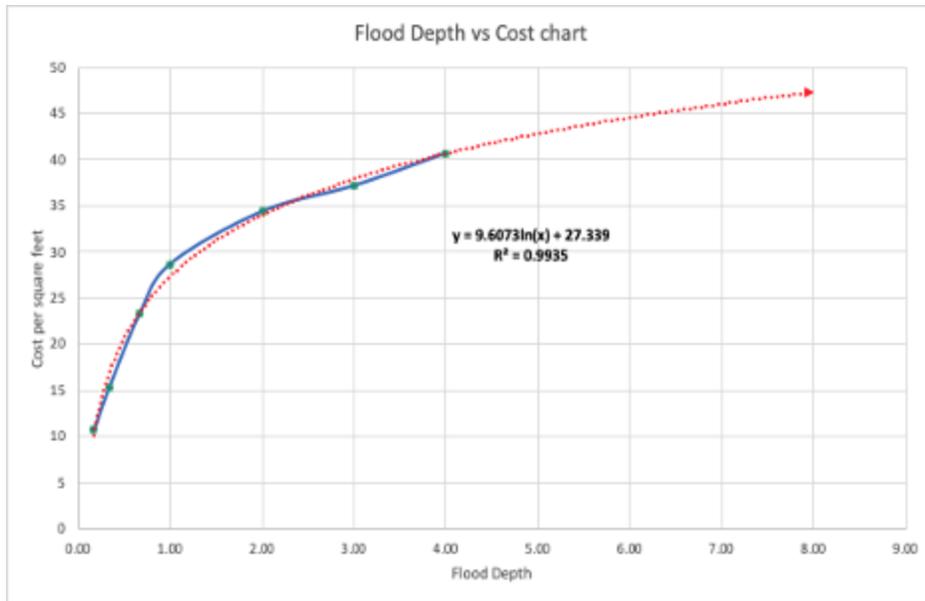

**Figure 4. Regression relationship derived from the FEMA floodsmart model.**

*Exploratory Data Analysis (EDA):*

The dataset was narrowed to only include observations where the currently assessed property (property plus land) value was greater than $10,000. In addition, the price per square foot had to be greater than one dollar. The reason for this subsetting is to only consider properties of value in the analysis. The dataset was further narrowed to only include observations with "base flood" values greater than zero because those are the properties that will be affected by flooding. The feature "area cost" was calculated by dividing features "shape area" and "land area" and the subsequent result was multiplied by the "current assessment" feature. Additionally, the observations were further subsetted to only include rows of data with an area cost higher than zero dollars in order to only consider properties of financial importance. Subsequently, outlier testing was conducted to exclude outlier observations. After subsetting the data, scatter plots were utilized in order to visualize the pairwise correlations between the features in the dataset.

## Results and Discussion

*Flood risk:*

The entire study area was analyzed for potential flooding with respect to the 1% annual chance flood elevation. The base flood areas of flooding and the associated depth of water was computed for each grid. Flood risk was also computed for varying sea level rise scenarios. These scenarios were then visualized in the map of the National Harbor. The flood hazard risk in the study area did not show much of a change (Fig 5). There seems to be a noticeable change between the scenarios towards the central part of the study area but a significant change in flood susceptibility was not noticeable at the mapped scale. But further inspection of the flooding areas indicate that the flooding indeed increased when going from the base level to sea-level scenarios. The increase in flooding area and the associated cost is shown in the table below (Table 1). The results show that sea level vulnerability is much more dependent on the terrain. As the terrain is less sloped, flood water will have an increased chance of propagating inland.

**Table 1. Changes in cost of flooding and total flooding area from base flood levels to increased sea levels.**

|  | Total flooding ($) | Total Area Flooded (ft$^2$) | Cost (% Δ) | Area (% Δ) |
|---|---|---|---|---|
| Base Flood | $106,302,284.38 | 49,073,440 |  |  |
| Sea Level (1ft) | $120,128,690.11 | 51,916,752 | 0.1301 | 5.79 |
| Sea Level (2ft) | $134,354,291.56 | 54,985,504 | 0.1338 | 6.25 |
| Sea Level (3ft) | $148,673,644.61 | 58,003,842 | 0.1347 | 6.15 |

Note: *The one-foot increments were assumed for the sea level rise scenarios to match up with the anticipated global sea level increase of one-foot increase per century.

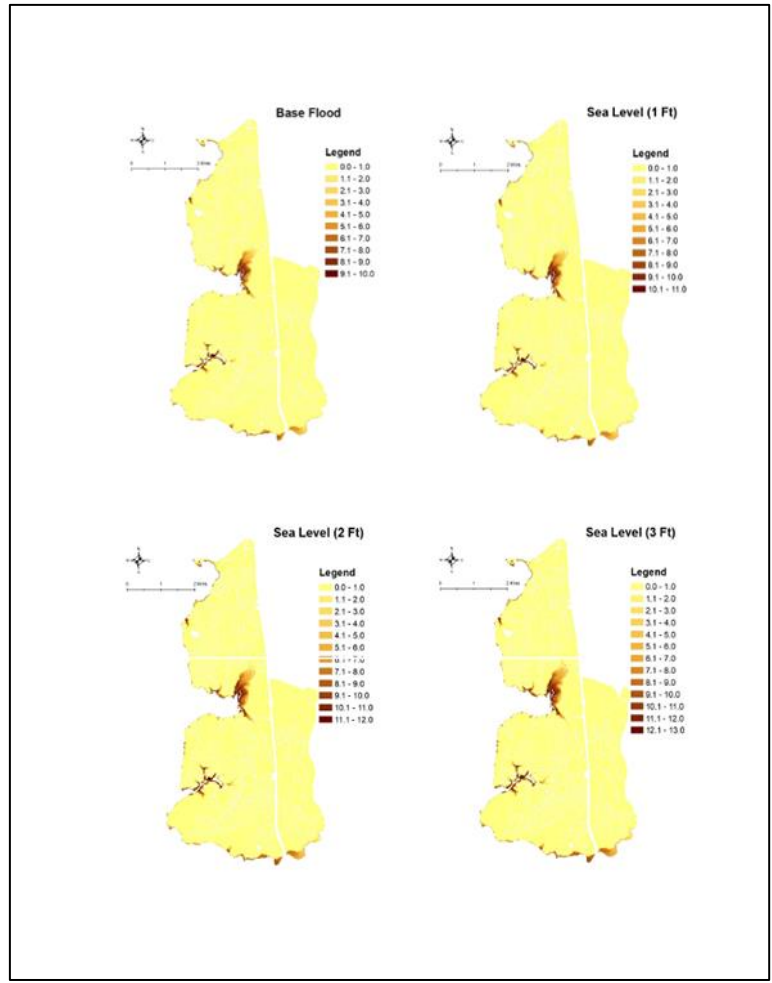

**Figure 5. Flooding extent for varying scenarios of sea level rise compared with base flood.**

*Area cost analysis:*

There was no strong relationship observed between area and current assessed value (Fig 6). In fact, the data points formed a funnel shaped distribution which is indicative of heteroskedasticity. This unequal scatter of data points demonstrates a non-constant variance in the dataset (James et al., 2013). Linear regression assumes homoscedasticity in the dataset (Yang et al., 2019) which means that the variance is constant. The non-linearity between the predictor (area) and the response (current assessed value) violates the linearity assumption of linear regression (James et al., 2013). Thus, the lack of homoscedasticity and linearity in the dataset makes it not ideal for fitting a linear regression model.

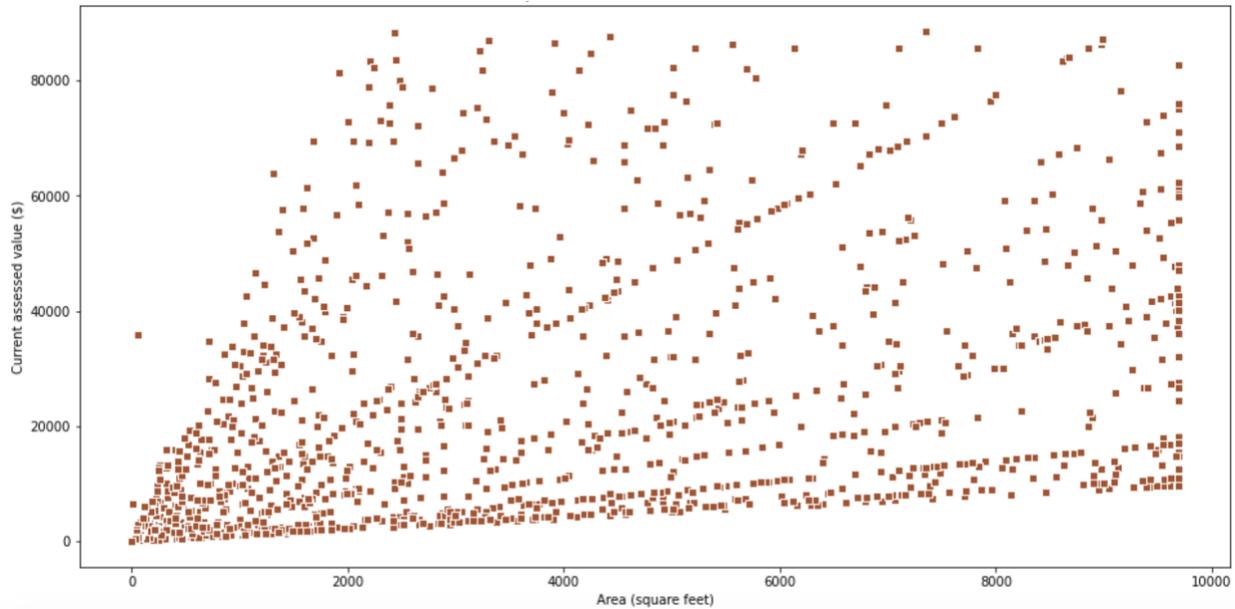

**Figure 6. Relationship between area (property plus land) and current assessed values.** Only observations with current assessed values greater than zero were included. Exploratory data analysis shows non-linearity in the dataset. The current assessed value corresponds to the feature 'area cost'.

## Conclusion

The study demonstrated that coastal flood risk increased with sea level rise even though the predicted level of impact is fairly insignificant for the study area. The level of impact from increased flooding is highly dependent on the location of the properties and other topographic information. Incorporating insurance quotes for individual properties would have made the analysis more complete and representative of our study area. With that said, the current model does not give us enough details about the cost of flooding as to how it relates to sea level rising. There is a need for further research to assess whether the dataset can be transformed to meet the homoscedasticity assumption of linear regression. In addition, it will be necessary to find new features with linear relationships between the cost of flooding to fulfill research objectives, and enrich the datasets with insurance premium and/or housing market data which can be potentially used to compare to the cost analysis.